\documentclass[12pt]{iopart}  

\usepackage{iopams}
\usepackage{graphicx}
\begin{document}

\title[S.V. Godambe et al.]{Very High Energy $\gamma$ -ray and Near Infrared observations of 1ES2344+514 during 2004-05}

\author{S. V. Godambe$^a$, R. C. Rannot$^a$,  K. S. Baliyan$^b$, A. K. Tickoo$^a$, S. Thoudam$^a$, V. K. Dhar$^a$, P. Chandra$^a$, K. K. Yadav$^a$, K. Venugopal$^a$, N. Bhatt$^a$, S. Bhattacharyya$^a$, K. Chanchalani$^a$, S. Ganesh$^b$, H. C. Goyal$^a$, U. C. Joshi$^b$, R. K. Kaul$^a$, M. Kothari$^a$, S. Kotwal$^a$, M.K. Koul$^a$, R. Koul$^a$, S. Sahaynathan$^a$, C. Shah$^b$, M. Sharma$^a$}

\address{(a) Astrophysical Sciences Division, Bhabha Atomic Research Centre, Trombay, Mumbai 400085, India}  
\address{(b) Physical Research Laboratory, Ahmedabad 380009, India}         
\ead{gsagar@barc.gov.in}

\begin{abstract}
We have observed the BL Lac object 1ES2344+514 (z = 0.044) in  Very High Energy
(VHE) gamma-ray  and near-infrared wavelength bands with TACTIC and MIRO
telescopes respectively. The observations  were made  from 18th October to 9th
December 2004 and 27th October 2005 to 1st January 2006. Detailed analysis of
the TACTIC data indicates absence of a statistically significant gamma-ray
signal both in overall data and on a nightly basis from the source
direction. We estimate an upper limit of I($\geq$1.5 TeV)$\leq 3.84 \times
10^{-12}$ photons cm$^{-2}$  s$^{-1}$ at a 3$\sigma$ confidence level on the
integrated $\gamma$-ray flux. In addition, we have also compared TACTIC TeV
light curves with those of the RXTE ASM (2-12keV) for the contemporary  period
and find  that there are no statistically significant increases  in  the  signal strengths  from the source in  both these energy regions. During 2004 IR
observations, 1ES2344+514 shows low level (~0.06 magnitude) day-to-day
variation in both, J \& H bands. However, during 2005 observation epoch, the
source brightens up by about 0.41 magnitude from its October 2005 level J magnitude
= 12.64  to J = 12.23  on December 6, 2005. It then fades by about
0.2 magnitude during 6 to 10 December, 2005. The variation is seen in both, J
\& H, bands simultaneously. The light travel time arguments suggest that
the emission region size is of the order of $10^{17}$ cms.

\end{abstract}

\section{Introduction}

Very High Energy  blazars belong to the the class of objects referred to as
radio-loud Active Galactic Nuclei (AGN) and are characterised by non-thermal
continuum spectrum, optical polarisation, flat radio spectrum and strong
variability in all frequency bands. Their broad band  Spectral Energy
Distribution (SED) which extends from radio up to the TeV range, consists
typically of two broad peaks \cite{bed97,Ulrich97,UrryPadovani95}. The low energy one peaks in the 0.1 - 100 keV domain and is commonly attributed to synchrotron emission
from ultra-relativistic leptons in the jet. The second component which is,
relative to the synchrotron emission, at high frequency is considered to be
due to the Inverse Compton (IC) scattering by the leptons of either
synchrotron photons (Synchrotron-Self-Compton process SSC) \cite{Bloom96,Konigl81,
Maraschi92, Sikora01}, or External Compton
(EC)\cite{Blandford95,Dermer92, Dermer93,Dermer97, Ghisellini96, Sikora94, Wagner95}. Alternative processes which are generally considered responsible for the second peak
are `hadronic' processes, 
which include interactions of a highly-relativistic jet
outflow with ambient matter  \cite{ Bednarek93,DarLaor97}, proton-induced cascades \cite{PIC} and synchrotron radiation due to protons \cite{Aharonian00,MP01}. 
Emission of TeV photons provides evidence for the presence
of particles at roughly the same energy if we assume IC radiation in the
Klein-Nishina regime \cite{BG70}. Therefore, it requires an extremely
efficient particle acceleration mechanism at work in the close environment of
the super massive black hole which forms the central engine of the AGN.

1ES2344+514 or QSO B2344+514 is a BL Lac type object (z $\sim$~ 0.044, 
l,b $\sim$112.89$^\circ$, -9.91$^\circ$) which has been the target of many studies. 
In the VHE $\gamma$-ray window, it was discovered by the Whipple Collaboration
in 1995 \cite{Catanese98}, wherein the evidence for emission came  mostly from an apparent flare on  December 20, 1995 and was subsequently confirmed independently by the HEGRA Collaboration \cite{Tluczykont03}. TeV spectra derived by the Whipple
group during the flaring state of the source in 1995 \cite{Schro05} are
steeper than for the brightest flare of Mrk421 and Mrk501, both located at
about 66$\%$ distance of  1ES2344+514.  The MAGIC collaboration have
reported VHE $\gamma$-ray signal in the energy range 140 GeV - 5.4 TeV during
the period August 3, 2005 - January 1, 2006 and found that the source was in a
low state \cite {magic06}. The differential spectra obtained by them shows a
steeper photon spectrum and six times lower flux level as compared to the
1995 flaring state. 
At X-ray energies (0.2-4 keV) it was detected by the 
Einstein Slew Survey \cite{Elvis92}. The source has also shown  a rapid x-ray
variability feature on a time scale of few hours during a  week-long campaign
in December 1996 using Beppo\textit{SAX} satellite in the 0.1 - 10 keV energy 
band when the source was in a  high state \cite{Giommi00}. 
In addition, a large decrease of 0.35 magnitude had been observed in the V
band during 3-17 January 2001 \cite{Xie02} indicating an optical variability. 

TeV Blazars are now known to undergo flaring episodes both at x-ray and TeV $\gamma$-ray energies  and the emissions   are generally correlated  albeit the  correlation appears to be fairly loose \cite{blaz05}.
By following the spectral evolution it is possible to test the models that predict
simultaneous flux changes in different parts of the spectrum, such as in the 
SSC model, where the x-ray and TeV
regions are expected to show correlated variations. Further, these objects are
characterised by variability features including short,  medium and long terms
variations and the  cross correlation of such features in different  wavelength bands may yield
new physics of these most violent objects. 
A strong correlation between GeV $\gamma$-ray and near 
infrared luminosities 
has been reported earlier\cite{Xie97} and it has been suggested that it may be
a common property of blazars. 

In this communication, we present TeV and near infrared results on 1ES2344+514      obtained during our 2004-05
observations with   TACTIC (TeV Atmospheric Cherenkov Telescope with Imaging Camera) gamma-ray and MIRO (Mount Abu Infrared Observatory)  telescopes respectively. 
In addition,  RXTE ASM\cite{ASM} x-ray data (2 - 12keV) of the contemporary period have  also been used to study the source behaviour in  three energy regimes. In the following sections we describe the  results  of these observations, including the data analysis procedure and follow it up  by a discussion and the main conclusions drawn from this study.

\section{TeV Observations}
\begin{table}
\caption {Observation log for 1ES2344+514 (ON Source Data)}
\begin{center}
\begin{tabular}{|c|c|c|c|c|}
\hline \hline
\hline \hline
\textbf{Year} &\textbf{Month} &\textbf{Observation} &\textbf{Total Obs.} &\textbf{Selected Obs.}\\
&       &\textbf{Dates} &\textbf{Time (hrs)} &\textbf{Time (hrs)}\\             
\hline \hline
\hline \hline
2004 &Oct.  &18-19   &3.21 &3.21   \\
2004 &Nov.        &3-5, 7-10, 15    &24.79      &11.72       \\
2004 &Dec.        &2-6, 8, 9        &19.14 &15.45       \\ 
\hline
\hline
\hline
2005 &Oct.        &27-29   &9.08   &9.08    \\ 
2005 &Nov.        &3, 20-22, 24-30 &19.41    &9.86      \\ 
2005 &Dec.        &1-5, 20-21, 24-30  &15.06   &12.52    \\ 
\hline 
\hline 
\hline 
\end{tabular}
\end{center}
\end{table}

\subsection {Experimental Setup}
  TACTIC gamma-ray telescope  is
located  at Mt. Abu (24.6$^\circ$ N, 72.7$^\circ$ E, 1300 m  asl), a hill
resort  in Western India. It uses a tessellated light-collector  
of  9.5 $m^2$ area which is configured as a quasi-parabolic surface, yielding
a measured spot-size of 0.3$^\circ$ for on-axis parallel rays. 
\begin{table}
\caption {Observation log for 1ES2344+514 (OFF Source Data)}
\begin{center}
\begin{tabular}{|c|c|c|c|c|}
\hline \hline
\hline \hline
\textbf{Year} &\textbf{Month} &\textbf{Observation}&\textbf{Total Obs.} &\textbf{Selected Obs.} \\
&       &\textbf{Dates} &\textbf{Time (hrs)} &\textbf{Time (hrs)} \\ 
\hline \hline
2004 &Dec.  &4-6, 8-9   &10.92 &10.92  \\
\hline
\hline
2005 &Dec.  &21-22, 24, 27-30   &8.84 &8.84  \\ 
\hline
\hline
2006 &Jan.        &1    &0.52  &0.52    \\
\hline
\hline
\hline
\end{tabular}
\end{center}
\end{table} 
The  PC-controlled 2-axes drive system of the telescope ensures a pointing /
tracking accuracy of better than $\pm$3 arc-mins. The pixel resolution of the
imaging camera is $\sim$ 0.31$^\circ$ throughout the camera FoV  of $\sim$
6$^\circ$ $\times$ 6$^\circ$. We have used the inner 225 pixels (15 $\times$
15 matrix) of the camera for the present studies.  The event-trigger
generation is based on the 3NCT (Nearest Neighbour Non-Collinear Triplets) logic for
the 2004 observations and  NNP ( Nearest Neighbour Pair) logic for the  2005-06
observations, demanding $\geq$ 8 photoelectrons  and $\geq$ 25 photoelectrons for the Triplet and
Pair pixels respectively, which participate in the trigger-generation.
Whenever the single's rate of one or more pixels goes outside the preset
operational band, it is automatically restored to within the prescribed range
by appropriately adjusting the pixel(s) high voltage(s). The resulting change
in the pixel(s) gain is monitored by repeatedly flashing a bright LED lamp,
improvised to produce a homogeneous light intensity  over the entire camera
\cite{Bhatt01}. From the logged digital counts (dc), the relative gains of all
the pixels are derived with respect to 4 'calibration' pixels for which the
high voltage is always kept fixed. In addition, these pixels, which are
located on the 4 edges of the camera are provided with $Am^{241}$-embedded
scintillator based optical flashers for on-line absolute calibration.  The
relative calibration and the sky pedestal data  of the camera pixels are
recorded several times in the course of observations for Cherenkov
image-cleaning and calibration purposes. The absolute occurrence time of each
individual event is recorded with a resolution of 1$\mu$s and an accuracy of a
few $\mu$s, using GPS provided reference time-markers for synchronisation of
the local observatory clock. The telescope is sensitive to $\gamma$-rays with
energies between approximately 1.5 TeV to 20 TeV  and can detect the Crab
Nebula at 5$\sigma$ significance level in 25 hours of on- source observation
\cite{Bhatt02}.

\subsection{Observations and Data Analysis} 
Observations on 1ES2344+514 were taken during  October 18 to December 9,
2004 and  October 27, 2005 to January 1, 2006, hereafter referred to as
spell 1 and spell 2 respectively, during cloudless and moonless nights. Total
duration of actual observations on the source  is 91.2 hours in a zenith  angle range of about 27 to 45 
degrees. Several standard data quality checks have been used to evaluate the
overall system behaviour and the general quality of the recorded data. These
include conformity of the prompt coincidence rates with the expected zenith
angle dependence, compatibility of the arrival times of prompt coincidence
events with Poissonian statistics and  the behaviour of the chance coincidence
rate with time. After applying these data quality checks we have selected 
good quality data sets of 60.15 hours as per the details given in Table 1. In
order to maximise the on-source observation time, we have used tracking
observation method\cite{Kerrick95}, for recording both the on and off-source
data, wherein background measurements are made from the candidate gamma-ray
source direction itself, thus eliminating the need for a dedicated off-source
observation run corresponding to each on-source run and ensuring that data are
taken under identical atmospheric conditions. However, independent off-source
data have also been collected for about 20.32 hours, details of which have
been given in  Table 2, to eliminate the possibility of  significant
systematic errors in the instrumentation and data analysis procedure.   
Accordingly, off -source data sets were also subjected to the same data
quality checks as were applied to the on-source data sets.

\begin{table}
 \caption{Imaging cut values used for analysis}
\begin{center}
\begin{tabular}{||c|c||}
\hline \hline
\textbf{Parameters}              &\textbf{Cut Values}\\
\hline \hline \hline
Length                  &0.11$^\circ$$\leq$ L$\leq$(0.155 + 0.0260*log(size))$^\circ$\\
\hline
Width                   &0.06$^\circ$$\leq$ W $\leq$ (0.080 + 0.01250*log(size))$^\circ$\\
\hline  
Distance                &0.4$^\circ$ $\leq$ D $\leq$ 1.3$^\circ$\\
\hline Size          &S $\geq$ 450 dc counts (6.5 digital counts =1.0 pe)\\
\hline
Alpha                   & $\alpha$ $\leq$ 18$^\circ$\\
\hline
\end{tabular} 
\end{center}
\end{table}
Detailed analysis of  imaging atmospheric Cherenkov telescope data involves
a number of steps including filtering of the light of  night sky background,
accounting for the differences in the relative gains of the PMTs, finding
Cherenkov image boundaries, image parameterization, event  classification, 
energy and direction determination etc. 
In the first step the pedestal mean value of a
Charge to Digital Converter (CDC), which  we estimate by artificially triggering the camera 2000 times, is subtracted from the CDC counts of each pixel of the camera.   In the second   step, the relative gain calibration related data is obtained by recording   2000 images triggered by an artificial pulsed light source (LED) in front of the camera surface. A light-diffusing medium in front of the pulsed light source
has also been used to ensure uniformity of the photon field across the camera
surface.  The refined calibration data  are then used to determine the
relative gains of each  pixel by comparing their mean signals with respect to
a reference pixel. In the next step, a picture threshold of 6.5 $\sigma$,
boundary threshold of 3$\sigma$ and an isolated picture threshold of 10
$\sigma$ have been used  to select maximum number of PMTs with signal and
boundaries of the event image, while  limiting the inclusion of pixels with
noise alone. After performing flat fielding and image cleaning  the next step
in the data analysis is the image parameterisation and event selection. 

Characterisation of the Cherenkov images formed at the focal plane  is 
done using the moment analysis of Hillas \cite{Hillas85} and various  image 
parameters, viz., Length (L), Width (W), Distance (D),  Alpha ($\alpha$) and
Size (S) are calculated for all the cleaned images. Based on  the dynamic
supercuts methodology \cite{Mohanty98} and  CORSIKA-based simulation studies
of the TACTIC \cite{Koul02}, $\gamma$-ray-like events are preferentially
picked from the overall data-base of the source by appropriately changing the
permitted L, W and D ranges as a function of S. The imaging  cuts used in the
analysis are listed in  Table 3.  

\begin{figure}
\begin{center}
\includegraphics[height=14cm,width=14cm,angle=0,clip]{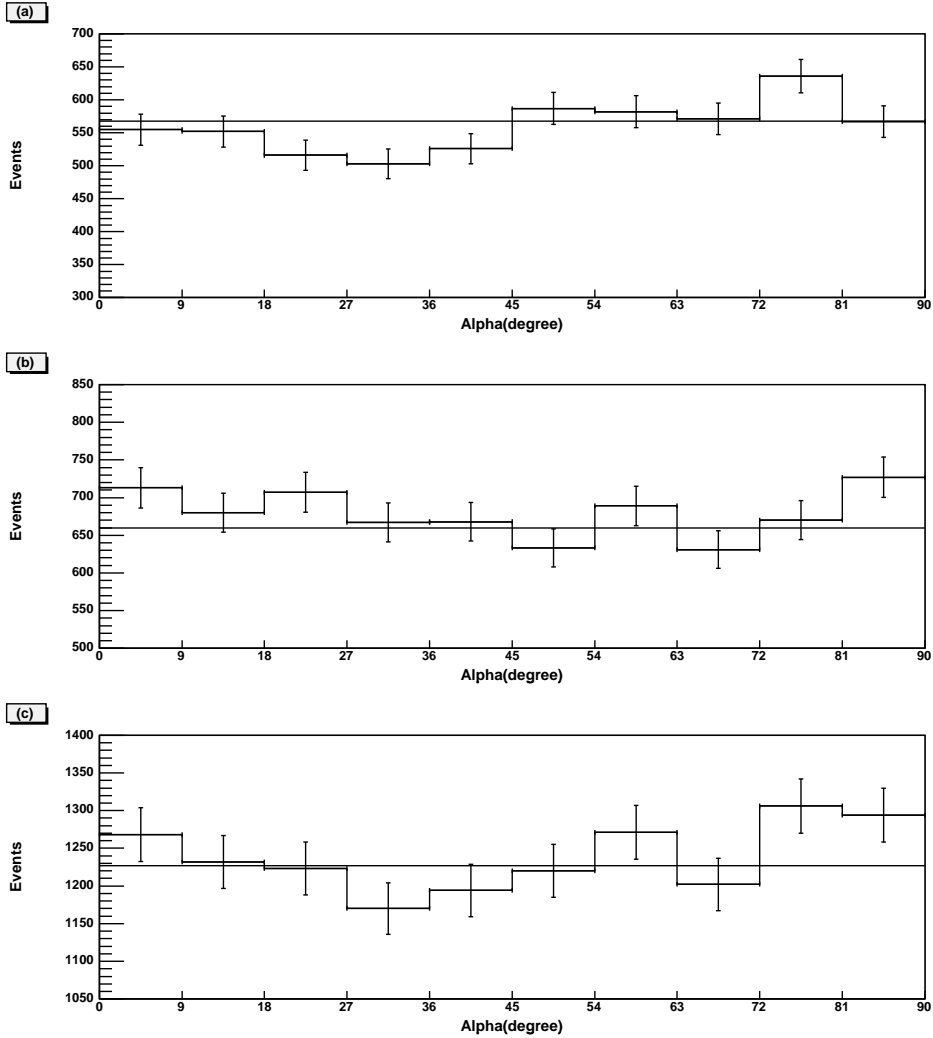}
\caption{Distribution of image parameter alpha from the on -source direction  (a) 2004 data (b) 2005 data and (c) for total data 2004-05.}
\end{center}
\end{figure}

\begin{figure}
\begin{center}
\includegraphics[height=14cm,width=14cm,angle=0,clip]{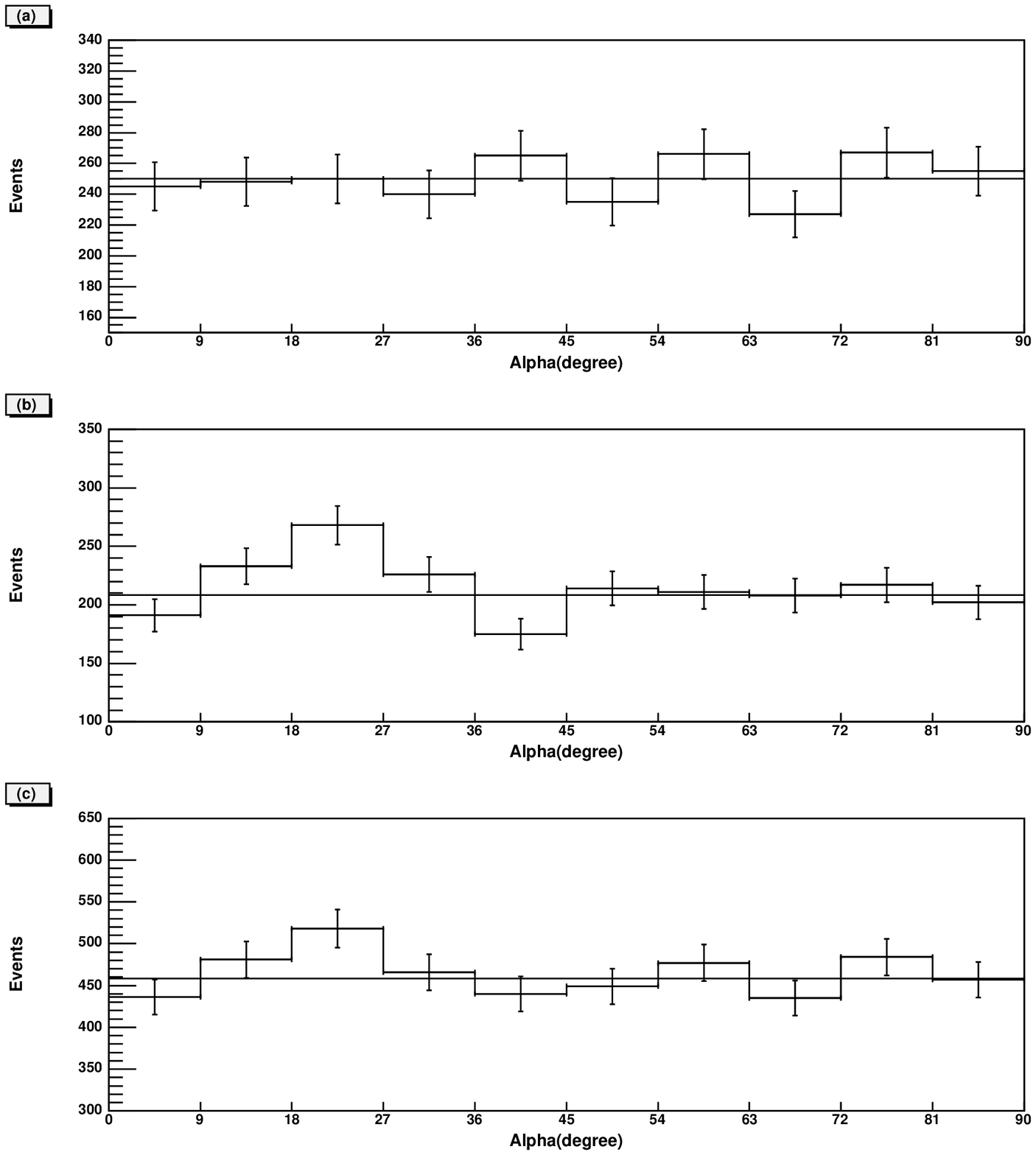}
\caption{Distribution of image parameter alpha from the off -source direction  (a) 2004 data (b) 2005 data and (c) for total data 2004-05.}
\end{center}
\end{figure}

\begin{figure}
\begin{center}
\includegraphics[height=9cm,width=12cm,angle=0,clip]{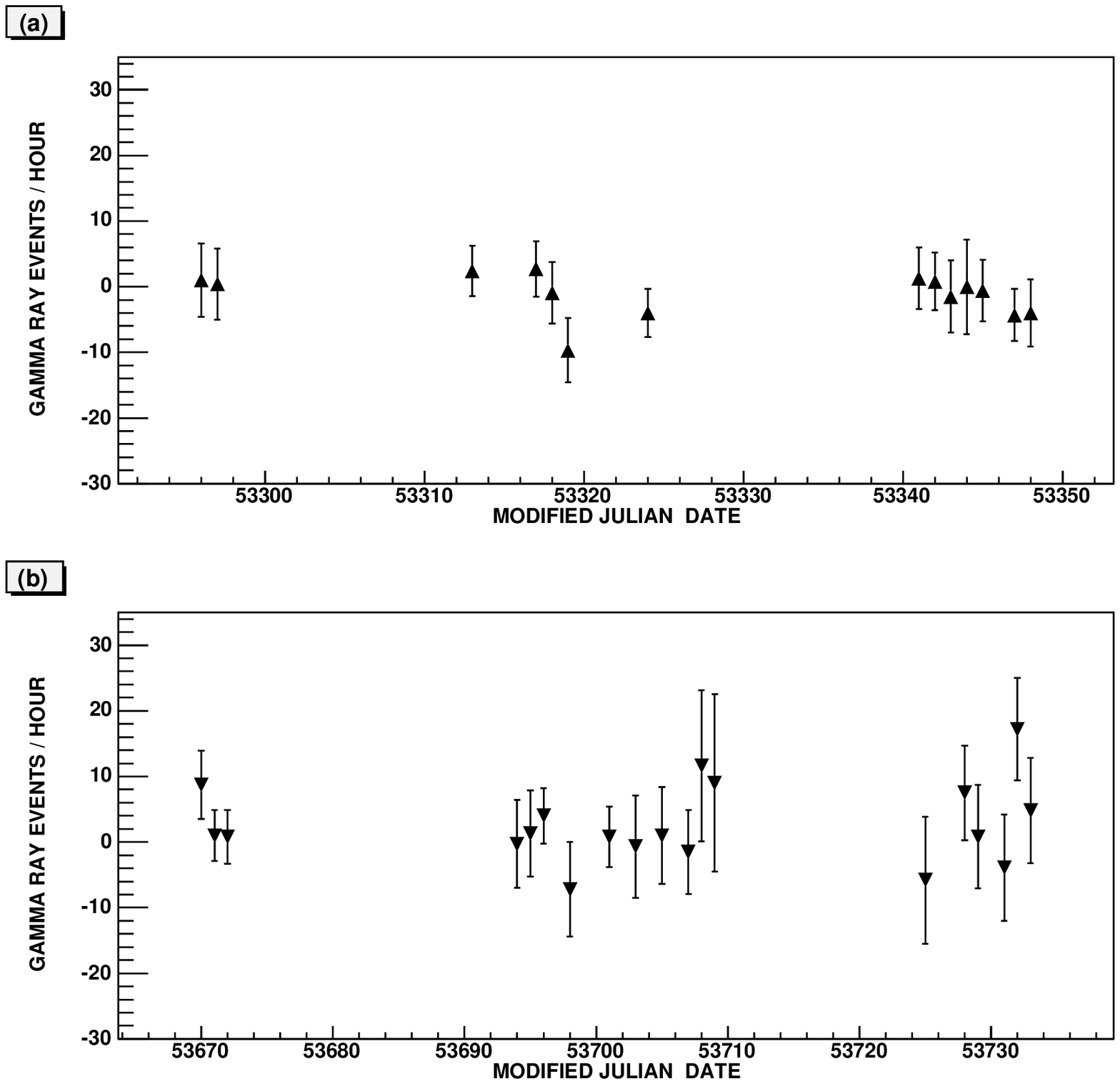}
\caption{TACTIC nightly gamma-ray rate variation for (a) spell 1 and (b) spell 2 data sets.}
\end{center}
\end{figure}

Under ideal conditions the alpha distribution is expected to be flat in the
absence of a gamma-ray signal from the source direction. From the TACTIC
simulation studies and  recent  observations of the Crab Nebula and Mrk421
 in flaring state \cite{kky2006}, we find that a TeV gamma-ray emitting source
shows an excess of events in the alpha range of 0$^\circ$ - 18$^\circ$ and  the
distribution is reasonably flat in the alpha range of 27$^\circ$ -
81$^\circ$. This flat distribution has been used for determining the
background level required for  estimating the on-source excess of events as is
discussed in \cite{kky2006}.

\subsection{Results of TeV Observations }
In Figure 1, we  show the derived alpha distributions after applying imaging
cuts as are given in  Table 3, for (a) spell 1 (b) spell 2 and (c) spell 1
and spell 2 together. It is clear that all the alpha plots are consistent with
flat distribution. None of the distributions of Figure 1 shows an excess  of
events in the  expected signal region of the alpha plot which is from
0$^\circ$ -18$^\circ$ \cite{kky2006} as has been mentioned earlier. Hence we
do not find  any evidence for a statistically significant TeV $\gamma$-ray
signal from the source direction   during  the two epochs  of TACTIC observations of
1ES2344+514. Excess or deficit obtained during spell 1 and 2 observations are
-28.0$\pm $ 38.5 and 73.6 $\pm $ 42.8 respectively, which are not
statistically significant. Accordingly we have derived an upper-limit on the
TeV gamma-ray flux at a 3 $\sigma$ confidence  level, of $<$5.08 $\times
10^{-12}$ photons cm$^{-2}$ s$^{-1}$ and $<$5.77 $\times 10^{-12}$ photons
cm$^{-2}$  s$^{-1}$ for spell 1 and 2 respectively, using the method of Helene
\cite{Helene83}. The  upper-limit on the integrated
$\gamma$-ray flux from the source direction above 1.5 TeV, estimated using the  data of spell 1 and 2 together ( excess  45.6 $\pm $ 47.5) is $<$ $3.84\times 10^{-12}$
photons cm$^{-2}$ s$^{-1}$ at the 3$\sigma$ confidence level.  We have
also carried out off -source data analysis using the same procedure and the
corresponding alpha plots are shown in  Figure 2 (a) spell 1 (b) spell 2 
and (c) both spells taken together, after applying various imaging  cuts, as
are given in  Table 3. These plots are also consistent with the expected distribution of the alpha parameter in the absence of a  gamma-ray signal.
In order to compare the TACTIC upperlimit with the MAGIC   group 
2005 detection of the source, we have  calculated  integral gamma-ray flux F($E_\gamma \geq$1.5 TeV ) =  $0.325 \times
10^{-12}$ photons cm$^{-2}$  s$^{-1}$  using their  differential  energy spectrum given in \cite{magic06,wagner06}, which is  clearly   much lower than the upperlimit obtained in this work. Therefore, the TACTIC upperlimit is in agreement with the MAGIC   group  detection in 2005. 

\begin{figure}

\begin{center}
\includegraphics[height=9cm,width=12cm,angle=0,clip]{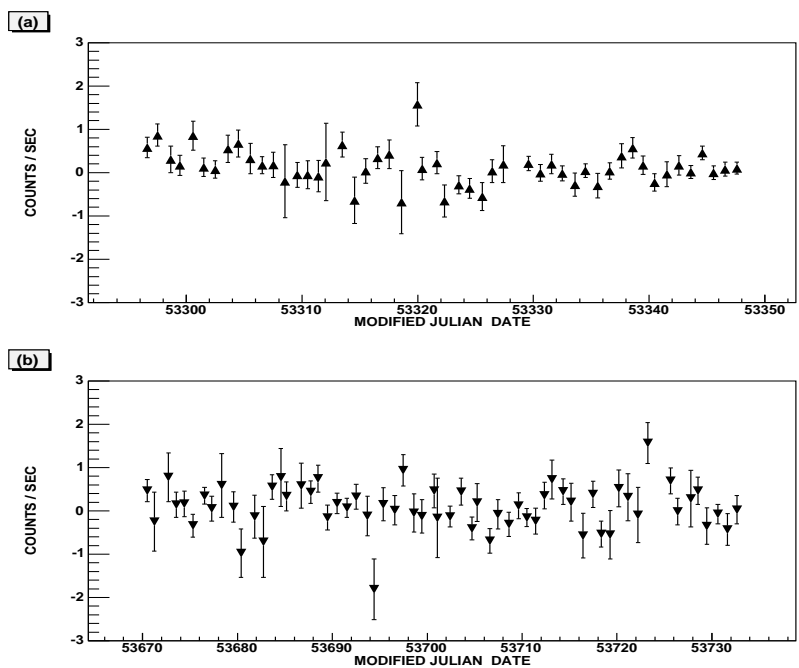}
\caption{RXTE/ASM (one day average  sum band intensity data)  light curves of 1ES2344+514 (a)  2004 and (b) 2005 data sets. }
\end{center}
\end{figure}
As mentioned earlier, the TeV blazars show a high flux- variability  on  time scales ranging from months to less than an hour. Interestingly, in   the Whipple discovery paper on the source \cite{Catanese98}, a flare of  VHE gamma-ray emission  (above 350 GeV) was  detected  on 20th Dec. 1995 during which a 6$\sigma$ excess was reported. In this work, we have also searched  for possible strong  TeV flaring episodes during  the epochs of our observations and accordingly divided the data
on nightly basis and repeated the analysis using the same methodology as mentioned earlier.  Figures  3 a and b show the day-to-day variations of the $\gamma$-ray rate ($\gamma$-rays/hour) for spell 1 and  spell 2 observations respectively. These two light curves are characterised with  reduced $\chi^2$ values of 0.42  and  0.75  respectively. The magnitude of an excess or deficit recorded on different nights is within $\pm$ 2 $\sigma$  level and hence indicates the absence of statistically significant variable TeV gamma-ray signal during the period of TACTIC observations. These results suggest that the source 1ES2344+514 was  possibly in a  quiescent state in the VHE gamma-ray range during  the two periods  of TACTIC observations. 

We have also shown for comparison in Figure 4, the corresponding light curve,
of the source in the x-ray energy range of 2-12 keV using the RXTE ASM one day average sum band intensity data
\cite{ASM}.  These two light curves for the period of spell 1 (Fig. 4a) and spell 2
(Fig. 4b)  are characterised by   reduced   $\chi^2$ values of 1.66 and 1.51
with respect to the average count levels of 0.12$\pm$0.03 and 0.11$\pm$0.04 respectively. In addition, we have also obtained the above said ASM light curves corresponding to the following three epochs: (1) Oct. -Dec. 1997, (2) Aug.- Nov. 1998 and (3) Sept. 2002, for which HEGRA group had reported 4.4 $\sigma$ detection in 72.5 hours, which mainly  resulted from their 1998 observations\cite{Aharonian04}.  The three  light curves are  characterised by  reduced $\chi^2$ values of 1.64, 1.45 and 2.11 with respect to the average count levels of 0.20$\pm$0.02 , 0.18$\pm$0.02 and  0.14$\pm$0.04 respectively thereby,   indicating   that the source was possibly in nearly similar emission state during the two TACTIC observation epochs (in terms of  variations on the time scale of a day) as it was during the epochs of the HEGRA  observations in the RXTE ASM energy range.

\begin{table}
\caption {Summary of TeV measurements of 1ES2344+514}
\begin{center}
\begin{tabular}{|c|c|c|c|c|}
\hline \hline
\hline \hline
\textbf{Spell}  &\textbf{N$_{on}$} &\textbf{N$_{off}$} &\textbf{Excess}  &\textbf{Upper Limit}\\
 & & & &photons cm$^{-2}$ s$^{-1}$ \\             
\hline \hline
\hline \hline
2004     &1107$\pm $ 33 &1135.$\pm $19  &-28.0$\pm $ 38.54      &       $<$5.08 $\times 10^{-12}$          \\
\hline
2005    &1393$\pm $37   &1319.33$\pm $ 20  &73.66$\pm $ 42.81   &       $<$5.77 $\times 10^{-12}$  \\
\hline 
2004-2005       &2500$\pm  $50  &2454.33$\pm $28   &45.66$\pm $ 57.6    &       $<$3.84 $\times$ 10$^{-12}$  \\
\hline \hline
\hline 
\hline 
\end{tabular}
\end{center}
\end{table}

\section{Near Infrared Observations and Results}

The Mount Abu Infrared Observatory  is located at Gurushikar, a few kms
away from the  TACTIC $\gamma$-ray telescope, at an altitude of 1680m. It
houses a 1.2m f/13 Cassegrain focus infrared telescope
\cite{desh95a,desh95b}. The observations in near infrared J and H bands were
taken  during: (1) November 18, 22,23,24 and December 5, 6 in 2004 and (2)
October 26,27 and December 6, 10,11 in  2005, using back end instrument NICMOS-3
near infrared detector array camera mounted  on the 1.2m IR
telescope. NICMOS-3 is a 256x256 pixels HgCdTe array detector, giving a FoV
of $\sim 4' \times 4'$ with a pixel resolution of 0.96 arc sec. The detector
is liquid nitrogen cooled to keep dark current low. In the near infrared
region the background is very high leading to detector saturation at longer
exposure times. To avoid detector saturation and improve upon the
signal-to-noise ratio in near infrared bands, a set of large number of frames
with short exposure times are taken at one position. Several such sets of
images are taken keeping the source at different locations on the detector
with 280 and 220 seconds of total integration times in J and  H bands,
respectively. At least two comparison stars are kept in each frame for
magnitude calibration. A large number of dark frames with same exposure times
as used in source observations were also taken every night. 
\begin{figure}
\begin{center}
\includegraphics[height=9cm,width=12cm,angle=0,clip]{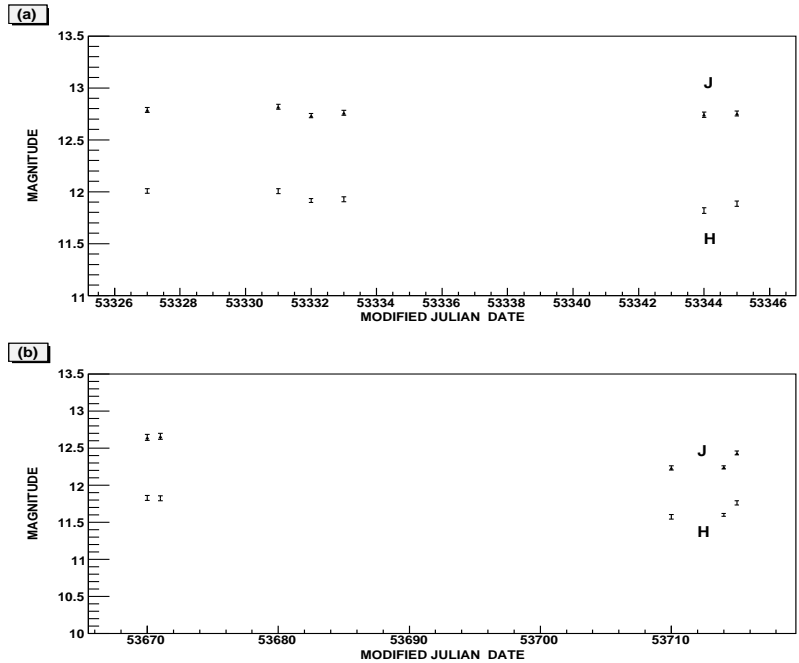}
\caption{ Near infrared J \& H band light curve for (a) 2004 and (b) 2005 observation
spells.}
\end{center}
\end{figure}

Data reduction and analysis was performed using IRAF and locally developed
scripts details of which are given elsewhere \cite{Baliyan05}. All the source
images in a set were combined and dark subtracted. To take care of background
sky, sky image was constructed from all the images taken during the night and
subtracted from the source images. The resultant 1ES2344+514 images were then
flat fielded and treated for bad pixels and cosmic ray correction. The final
images containing the source, 1ES2344+514, and calibration stars were
subjected to aperture photometry keeping the aperture size same for the source
and the comparison stars. The instrumental magnitudes thus obtained were used
for differential photometry light curves. The source magnitudes were corrected
using standard magnitude values of the comparison stars and these corrected J
and H band magnitudes are plotted in Figure 5 as a function of time in MJD. It
is clear from the figure  that during 2004 observations, 1ES2344+514 shows low
level (~0.06 magnitude) day-to-day variation in both, J \& H bands. The J band
magnitude of the source remains around 12.85 which is about the average value
for the source. The source, therefore, does not show any flaring/outburst
activity in this energy regime during this period. It should be noted that the
host galaxy of the 1ES2344+514 is relatively bright and stellar light from the 
host is capable of affecting the measurement of nuclear emissions,
particularly when the seeing conditions are not good \cite{Cellone2000}. Under such observing conditions, while making observations in optical and near infrared, it is
necessary to keep the aperture size fairly large in order to minimise the
effect of host emission on the intrinsic variations in the nuclear flux. On
the other hand, too large an aperture allows dilution of the nuclear emission
and dampens amplitude of its intrinsic variation. Both these factors decide
the size of aperture used in the photometric observations when the host is
bright. We have taken care of these aspects in the present study by following the recommendation given in  \cite{Cellone2000}. In our case, the seeing FWHM
varied from about 2.5$^{\prime\prime}$ to  3.3$^{\prime\prime}$ (in the worst case) and therefore to optimize the photometric results, we have used 8 pixels as photometric aperture radius. The choice enables us to achieve good S/N ratio while
keeping any spurious variation in source brightness, caused by seeing change
and host galaxy contribution, below 0.01 mag. However, during most of the
nights observing conditions were very good, photometric or close to
photometric. We exercised extra precaution whenever we noticed variation in
seeing. In particular, during the October 2005 epoch when the sky conditions were not
stable leading to poor seeing, we used several standard stars to calibrate the
source magnitude apart from the measures mentioned above. Stars were chosen to have about same magnitude as the source as recommended by \cite{Cellone07,Howel1988}. 

In the 2005 observation epoch, the source is seen in fainter (J magnitude
$\sim$ 12.6) state during October 26- 27 (MJD 53670- 53671), brightening up by 0.41 magnitude on  December 6, 2005 (MJD 53710). It then fades by about 0.2 magnitude during 6 to 10 December, 2005. However, intra-night variations appear to be  within 1 $\sigma$ level. It shows similar behaviour in both, J and H, bands. Since we do not
have a continuous coverage of the source during October - December, 2005, it
is not possible for us to state whether 1ES2344+514 had several flares or a
slow increase in flux during the brightening phase. It would be very
interesting to know  the  behaviour of the source in radio, optical and x-ray
regions in order to infer the nature of  near infrared emission - whether it
is thermal or mostly non-thermal. The RXTE ASM(2-12 keV) x-ray data used here
does not indicate any significant variation during this period. We are not
aware of any observations in radio or optical region which could throw more
light on the nature of the source during this period. 

\begin{figure}
\begin{center}
\includegraphics[height=9cm,width=12cm,angle=0,clip]{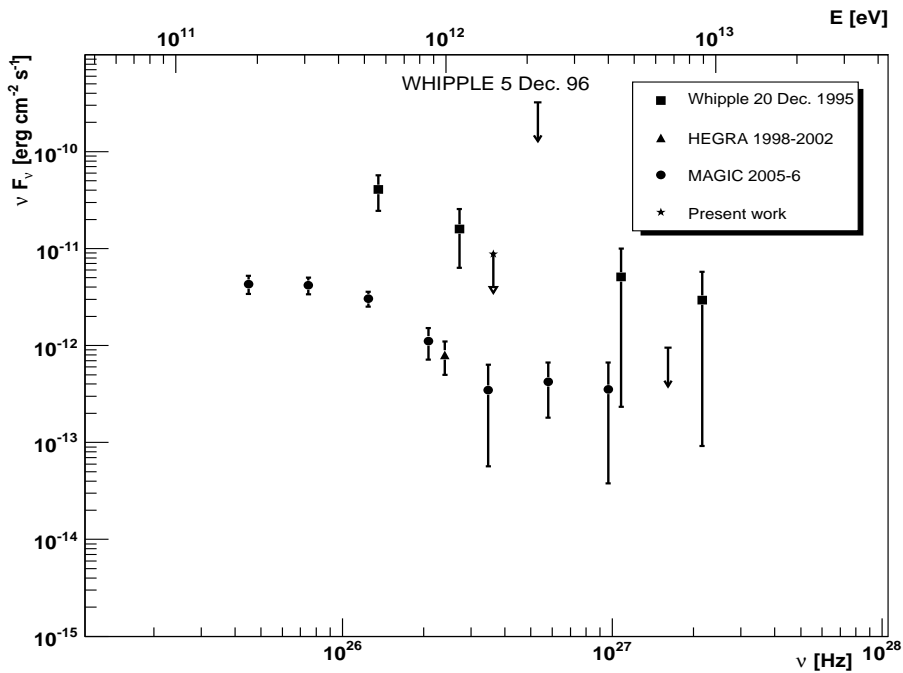}
\caption{  TACTIC upperlimit derived using 2004-05 on-source data for 60.15 hours  has been shown along with  other VHE results from Whipple \cite{Schro05}, HEGRA \cite{Aharonian04} and MAGIC \cite{magic06,wagner06} groups  on the blazar 1ES2344+514.}
\end{center}
\end{figure}

\section{Discussion and Conclusions}
We have  studied  the BL Lac object 1ES2344+514  in  
VHE gamma-ray  and near infrared wavelength bands with TACTIC and MIRO
telescopes respectively during 2004-2005. We do not find any evidence for the presence of a statistically significant VHE gamma-ray
signal, both in the overall data and on a nightly basis from the source
above the threshold energy of 1.5 TeV. An upper limit of I($\geq$1.5 TeV)$\leq 3.84 \times10^{-12}$ photons cm$^{-2}$  s$^{-1}$ has been obtained at a 3$\sigma$ confidence level on the integrated $\gamma$-ray flux and we conclude that the source was below the  upper limit flux during the periods of these observations. 
The derived upperlimit has  been shown in Figure 6 for  comparison with the  other VHE results from  Whipple \cite{Schro05}, HEGRA \cite{Aharonian04} and MAGIC \cite{magic06,wagner06} groups  on the same blazar 1ES2344+514. As is clear from  this figure   the TACTIC upperlimit is not in  conflict  with the MAGIC  \cite{magic06,wagner06} group detection  of the source in 2005, as it is placed at  I($E_\gamma \geq$1.5 TeV)$\leq 3.84 \times10^{-12}$ photons cm$^{-2}$  s$^{-1}$  whereas the MAGIC detected integral flux F($E_\gamma \geq$1.5 TeV ) =  $0.325 \times
10^{-12}$ photons cm$^{-2}$  s$^{-1}$ is much lower than the TACTIC upperlimit. 
In the RXTE ASM energy domain (2-12 keV) too, the light curves obtained from the RXTE site \cite{ASM}, which we have shown in  Figure 4, do not show any statistically significant evidence for  variations in the day averaged sum band intensity counts. 

Our observational results in the  near infrared wavelength bands are, however  mixed, wherein the source 1ES2344+514 shows low level (~0.06 magnitude) day-to-day variation in both, J \& H bands  during 2004 observations. Whereas, during 2005  observations in the same bands, the source brightens up by about 0.41 magnitude from its October 2005 level J magnitude = 12.64  to J = 12.23 magnitude on December 6, 2005. It then fades by about 0.2 magnitude during 6 to 10 December, 2005. The variation is seen in both, J \& H, bands simultaneously. From these near infrared observations, we conclude that the source has possibly undergone a brightening phase during  2005 MIRO observations. 

Further, the flux variations  within a timescale of a day in the RXTE ASM and TACTIC energy ranges are low, when the source brightens by 0.41 magnitude in low energy near
infrared region over a period of about 42 days or fades by 0.2 magnitude over
a period of about 5 days. By considering the light travel time arguments and
taking the fastest source variation in this study (0.2  magnitude in 5-days),
the size of the near infrared emission region is about 5-light days
(1.08$\times 10^{17}$ cm). Here, we have taken into account the Doppler factor of the jet as 8.4, obtained by \cite {magic06} for the low state of the source and assuming that the variable near infrared emission is likely  produced in the relativistic jet.

This source is dimmer than Mrk 421 and Mrk 501 at every wavelength  so it is
perhaps not surprising that the VHE emission of 1ES2344+514 appears to be
weaker on average. Long term simultaneous multi-wavelength observations, with
improved sensitivity particularly in TeV domain, are required to understand
the emission mechanisms. 
   
\section{ Acknowledgements}
The authors would like to convey their gratitude to all the concerned colleagues of the Astrophysical Sciences Division for their contributions towards the instrumentation and observation aspects of the TACTIC telescope. The  valuable suggestions from anonymous referees  are gratefully acknowledged. Finally, we are thankful to Dr. R. M. Wagner of Max-Planck-Institut f\"ur Physik, D-80805 M\"unchen, Germany,  for providing  useful inputs with respect to  earlier 1ES2344+514 work.

\end{document}